\documentclass{acm_proc_article-sp}
\usepackage{url}
\usepackage{color}
\usepackage{pbox}

\usepackage{setspace}

\usepackage{etoolbox}
\usepackage{hyperref}

\makeatletter
\patchcmd{\maketitle}{\@copyrightspace}{}{}{}
\makeatother

\begin{document}

\title{Fast filtering and animation of large dynamic networks}

\numberofauthors{3} 
\author{
\alignauthor
Przemyslaw A. Grabowicz\\
       \affaddr{IFISC, UIB, Spain}\\
       \affaddr{MPI-SWS, Germany}\\
\alignauthor
Luca Maria Aiello\\
       \affaddr{Yahoo! Research, Spain}\\
\alignauthor Filippo Menczer\\
       \affaddr{CNetS, Indiana University, USA}\\
}

\maketitle

\begin{abstract}
Detecting and visualizing what are the most relevant changes in an evolving network is an open challenge in several domains. We present a fast algorithm that filters subsets of the strongest nodes and edges representing an evolving weighted graph and visualize it by either creating a movie, or by streaming it to an interactive network visualization tool. The algorithm is an approximation of exponential sliding time-window that scales linearly with the number of interactions. We compare the algorithm against rectangular and exponential sliding time-window methods. Our network filtering algorithm: i) captures persistent trends in the structure of dynamic weighted networks, ii) smoothens transitions between the snapshots of dynamic network, and iii) uses limited memory and processor time. The algorithm is publicly available as open-source software.
\end{abstract}

\section{Introduction}

Network visualization is widely adopted to make sense of, and gain insight from, complex and large interaction data. These visualizations are typically static, and incapable to deal with quickly changing networks. Dynamic graphs, where nodes and edges churn and change over time, can be effective means of visualizing evolving networked systems such as social media, similarity graphs, or interaction networks between real world entities.
The recent availability of live data streams from online social media motivated the development of interfaces to process and visualize evolving graphs. Dynamic visualization is supported by several tools~\cite{erten04graphael, gleamviz, bastian09gephi, dutot07graphstream}. In particular, Gephi~\cite{bastian09gephi} supports graph streaming with a dedicated API based on JSON events and enables the association of timestamps to each graph component.

While there is some literature on dynamic layout of graphs~\cite{brandes05dynamic,brandes07dynamic,brandes11visualization}, not much work has been done so far about developing information filtering techniques for dynamic visualization of large and quickly changing networks. Yet, for large networks in which the rate of structural changes in time could be very high, the task of determining the nodes and edges that can represent and transmit the salient structural properties of the network at a certain time is crucial to produce meaningful visualizations of the graph evolution.


We contribute to filling this gap by presenting a new graph filtering and visualization tool called \texttt{fastviz} that processes a chronological sequence of weighted interactions between the graph nodes and dynamically filters the most relevant parts of the network to visualize. Our algorithm:
\begin{itemize}
	\item captures persistent trends in structural properties of dynamic networks, while removing no longer relevant portions of the networks and emphasizing old nodes and links that show fresh activity;
	\item smoothens transitions between the snapshots of a dynamic network by leveraging short-term and \text{long-term} node activity;
	\item uses limited memory and processor time and is fast enough to be applied to large live data streams and visualize their representation in the form of a network.
\end{itemize}

The reminder of this paper is structured as follows. First, we introduce related studies in Section 2. Next, we introduce the \texttt{fastviz} filtering method for dynamic networks in Section 3. We compare this method against rectangular and exponential sliding time-window approaches and show what are the advantages of our method. Finally, we present visualizations created with our filtering methods for four different real datasets in Section 4, and conclude the study.

\section{Related Work}

Graph drawing~\cite{kamada89visualizing,tollis99graph} is a branch of information visualization that has acquired great importance in complex systems analysis. A good pictorial representation of a graph can highlight its most important structural components, logically partition its different regions, and point out the most central nodes and the edges on which the information flows more frequently or quickly.
The rapid development of computer-aided visualization tools and the refinement of graph layout algorithms~\cite{kamada89algorithm,fruchterman91graph,gansner98improved,hu05efficient} allowed increasingly higher-quality visualizations of large graphs~\cite{herman00graph}.
As a result, many open tools for static graph analysis and visualization have been developed in the last decade. Among the best known we mention Walrus~\cite{munzner00interactive}, Pajek~\cite{batagelj02pajek,denovy05exploratory}, Visone~\cite{brandes03visone}, GUESS~\cite{adar06guess}, Networkbench~\cite{networkworkbench}, NodeXL~\cite{smith09analyzing}, and Tulip~\cite{auber12tulip}. Studies about comparisons of different tools have also been published recently~\cite{ahn11temporal}.

The interest in depicting the shape of online social networks~\cite{heer05vizster,falkowski06mining} and the availability of live data streams from online social media motivated the development of tools for animated visualizations of \textit{dynamic graphs}~\cite{demetrescu10dynamic}, in \textit{offline} contexts, where temporal graph evolution is known in advance, as well as in \textit{online} scenarios, where the graph updates are received in a streaming fashion~\cite{brandes05dynamic}. Several tools supporting dynamics visualization emerged, which include GraphAEL~\cite{erten04graphael}, GleamViz~\cite{gleamviz}, Gephi~\cite{bastian09gephi}, and GraphStream \cite{dutot07graphstream}. Despite static visualizations based on time-windows~\cite{ahn11temporal}, alluvial diagrams~\cite{rosvall10mapping}, or matrices~\cite{dyi10timematrix,stein10pixel,gove11netvisia} have been explored as solutions to capture the graph evolution, dynamic graph drawing remains the technique that has attracted more interest in the research community so far. Compared to static visualizations, dynamic animations present additional challenges: user studies have shown that they can be perceived as harder to parse visually, even though they have the potential to be more informative and engaging~\cite{farrugia11effective}.

As a result, a large corpus of work about the theoretical concepts on good visualization practices, especially for dynamic graphs, has been produced in the last two decades. Besides the work done in defining efficient update operations on graphs~\cite{ramalingam96computational,henzinger99randomized}, several principles about good graph visualizations have been proposed and explored in different studies. Friedrich and Eades~\cite{friedrich02graph} defined high-level guidelines for a good visualization of graph evolution with animations, including uniform, smooth and symmetrical movement of graph elements, with minimization of edge crossings and overviewing some techniques that make the visualization more enjoyable, such as fadeout deletion of nodes. Graph \textit{readability} has been measured in user studies in relation to several tasks~\cite{purchase97aesthetics,huang06people,huang08beyond}; the experimental findings highlight the importance of visualization criteria such as minimizing bends and edge crossings and maximizing cluster separation in facilitating the viewer's interpretation and understanding of the graph. A general concept that has been studied for long in relation to the quality of dynamic graph visualization is the \textit{mental map}~\cite{eades91preserving,misue95layout,freire06preserving} that the viewer has of the graph structure. In practical terms, the placement of existing nodes and edges should change as little as possible when a change is made to the graph~\cite{coleman96aestetics}, under the hypothesis that if the mental map is preserved the parsing of the visual information is faster and more accurate. More recent work~\cite{archambault13map} has reappraised the importance of the mental map in the comprehension of a dynamic graph series, while identifying some cases in which it may help~\cite{archambault12mental,archambault13mental} (e.g., memorability of the graph evolution, following long paths, recognition of recurrent patterns, tracking a large number of moving objects).

More in general, there are several open fronts in empirical research in graph visualization to identify the impact of certain factors on the quality of the animation (e.g., speed~\cite{ghani12perception}, interactivity~\cite{archambault08grouseflocks}). An extensive overview of this aspect has been conducted recently by Kriglstein et al.~\cite{kriglstein12information}. Methods to preserve the stability of nodes and the consistency of the network structure leveraging hierarchical organization on nodes have been proposed~\cite{north95incremental,north01online,archambault07visualization,archambault09structural}. User studies have shown that hierarchical approaches that collapse several nodes in larger meta-nodes can improve graph readability in cases of high edge density~\cite{archambault10readability}. The graph layout also has a significant impact on the readability of graphs~\cite{blythe95effect}. Some work has been done to adapt spectral and force-directed graph layouts~\cite{brandes01drawing} to \textit{incremental layouts} that recompute the position of nodes at time $t$ based on the previous positions at time $t$-$1$ minimizing displacement of vertices~\cite{branke01drawing,dihel02graphs,brandes05dynamic,frishman08online} or to propose new ``stress-minimization'' strategies to map the changes in the graph~\cite{brandes11visualization}.

Although much exploration has been done in the visualization principles to achieve highly-readable animations, two aspects have been overlooked so far.

First, not many techniques to extract and visualize the most relevant information from \textit{very large} graphs have been studied yet. Graph decomposition has been used in a static context to increase the readability of the network by splitting it into modules to be visualized separately~\cite{rodriguez11group}, while sliding time-windows have been employed to discard older nodes and edges in visualization of graph evolution~\cite{dynes04temporal}. A hierarchical organization of nodes according to some authority or centrality measure allows to visualize the graph at different levels of details, eliminating the need to display all nodes and edges at once~\cite{kumar06visual}. Some work has been done about interactive exploration by blending different visualization paradigms~\cite{hadlak11insitu} and time-varying clustering~\cite{sallaberry12clustering}. Indices to measure the relevance of events in a dynamic graph at both node and community level have also been proposed~\cite{asur07event}, even if they have not been applied to any graph animation task. Yet, none of these techniques has been tested on very large data and none of the modern visualization tools provide features for the detection of the most relevant components of a graph at a given time. On the other hand, quantitative studies on the characterization of temporal networks~\cite{clauser07persistence,cattuto10dynamics,krings12effects} have been conducted, but with no direct connection with the dynamic visualization task.

Last, the visualization of large graphs in an online scenario, where node and edge updates are received in a live stream, and the related practical implications of dynamic visualizations, have rarely been considered. In this context, just some exploratory work has been carried out about information selection techniques for dynamic graph visualization, including solutions based on temporal decay of nodes and edges~\cite{dynes04temporal}, node clustering~\cite{rodriguez11group}, and centrality indices~\cite{kumar06visual,asur07event}.

\section{Network filtering}

We introduce the \texttt{fastviz} algorithm that takes in input a chronological stream of interactions between nodes (i.e., network edges) and converts it into a set of graph updates that account only for the most relevant part of the network. The algorithm has two stages: buffering of filtered network and generation of differential updates for the visualization (see Figure~1). The algorithm stores and visualizes the nodes with the highest strengths, i.e., the highest sum of weights of their connections.

\begin{figure}[tbp]
\begin{center}
\includegraphics[width=5.5cm]{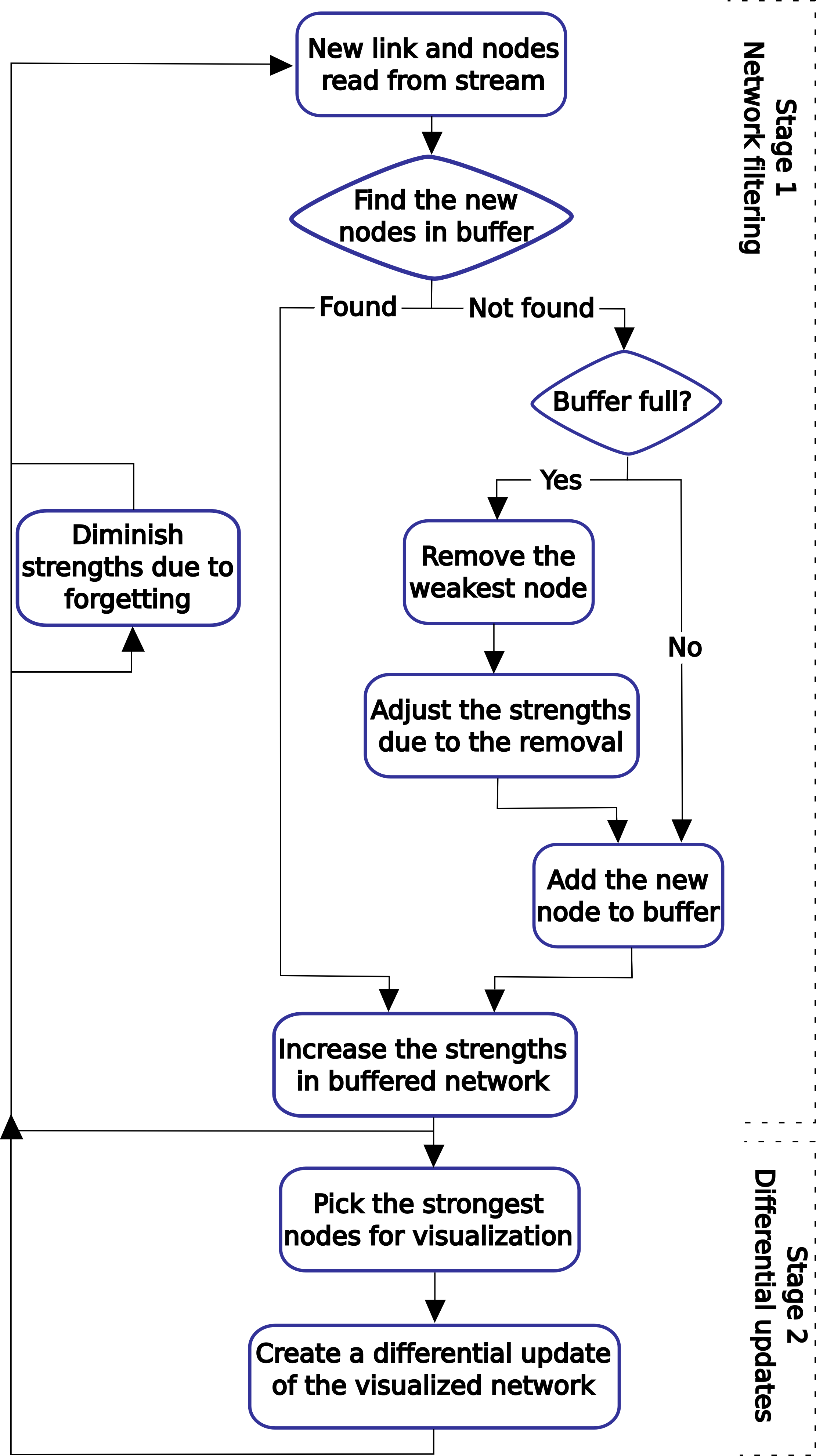}
\caption{The diagram illustrating the main components of the algorithm.}
\end{center}
\end{figure}

\subsection*{Input}

The data taken as input is an ordered chronological sequence of interactions between nodes. The interactions can be either pairwise or cliques of interacting nodes. For instance, the following input:
\begin{eqnarray*}
\left\langle  t_i, n_1, ..., n_m, w_i \right\rangle
\end{eqnarray*}
represents the occurrence of interactions between nodes $n_1, ..., n_m$ of weight $w_i$ at epoch time $t_i$. Entries with more than two nodes are interpreted as interactions happening between each pair of members of the clique with the respective weight. Multiple interactions between the same pair of nodes sum up by adding up their corresponding weights. The advantage of the clique-wise format over the pairwise format is that the size of input files is smaller.

\subsection*{Filtering criterion}

In the first stage of the algorithm, at most $N_\text{b}$ nodes with the highest strengths are saved in the buffer together with the interactions among them. The strength $S_i$ of a node $i$ is a sum of weights of all connections of that node, i.e., $S_i=\sum_j w_{ij}$, where $w_{ij}$ is the weight of an undirected connection between nodes $i$ and $j$. Whenever a new node, which does not appear in the buffer yet, is read from the input, it replaces the node in the buffer with the lowest value of the strength. If an incoming input involves a node that is already in the buffer, then the strength of the node is increased by the weight of the incoming connection.
To emphasize the most recent events and penalize stale ones, a forgetting mechanism that decreases the strengths of all nodes and weights of all edges is run periodically every time period $T_\text{f}$ by multiplying their current values by a forgetting factor $0\leq C_\text{f}<1$. This process leads to the removal of old inactive nodes having low strength and storage of old nodes with fresh activity and high strength.

Note that the forgetting mechanism corresponds to a sliding time-window with exponential decay. The decay determines the weighting of the past interactions in the sliding time-window aggregation of a dynamic network. Standard rectangular sliding time-window aggregates all past events within the width $T_\text{tw}$ of the time-window weighting them equally. In contrast, in \texttt{fastviz} and in the sliding time-window with an exponential decay the weighting decreases exponentially (see Figure~2).\footnote{Under a set of assumptions one can calculate how much time will a given node stay in the buffered network. Let us assume that at the time $t_n$ the strength of a node $n$ is $S_n(t_n)$, that this strength will not be increased after time $t_n$, that the next forgetting will happen in $T_\text{f}$ time, and that the strength of the weakest buffered node $S_\text{w}<S_n(t_n)$ is constant over time. Under these assumptions, the node $n$ will stay buffered for time $t-t_n>\frac{\log(S_\text{w}/S_n(t_n))}{\log(C_\text{f})} \, T_\text{f}$.}
Such exponential decay has two advantages over a standard rectangular sliding time-window approach. First, it gives more importance both to the most recent and to the oldest connections, while giving less importance to the middle-aged interactions. Second, it produces a dynamic network in which changes are smoother due to the balanced weighting of old and new connections. Finally, instead of using the sliding time-window with exponential decay, we introduced the \texttt{fastviz} algorithm to limit the computational complexity of network filtering. In principle, time-window methods do not introduce such a bound. We explore and confirm these points in the following subsections using real dynamic networks.

\begin{figure}[tbp]
\begin{center}
\includegraphics[width=8.5cm]{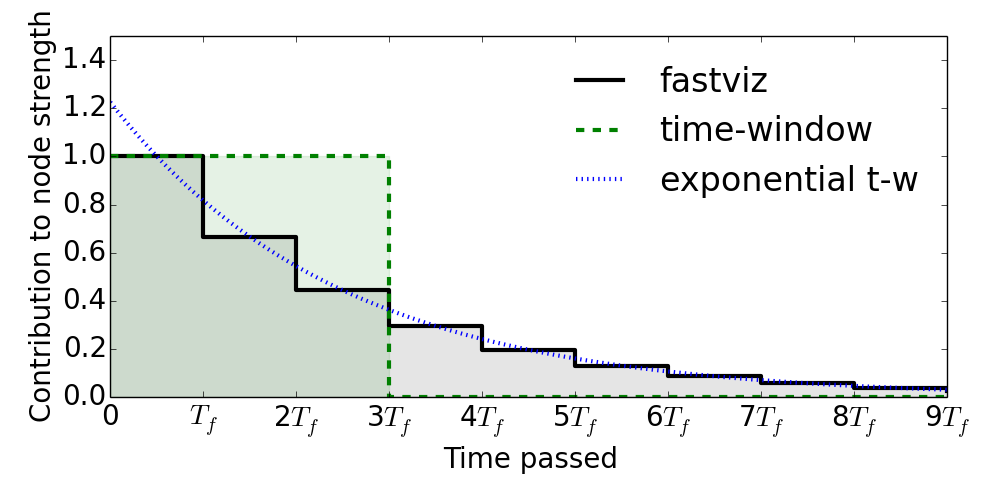}
\caption{The aggregating curves for \texttt{fastviz} (black line), rectangular sliding time-window (green dashed), and exponential time-window (blue dotted). The steps of the \texttt{fastviz} method correspond to consecutive multiplications by the forgetting factor $C_\text{f}=2/3$ performed after each forgetting period $T_\text{f}$. The rectangular time-window width is set to $T_\text{tw}=3\, T_\text{f}$. The exponent of the exponentially decaying time-window corresponds to the forgetting factor of \texttt{fastviz}. For these values of the parameters, areas under the aggregating curves of both methods are approximately equal, according to Equation \ref{eq:params}.}
\end{center}
\end{figure}

\subsection*{Filtering criterion versus rectangular and exponential sliding time-windows}

Comparison of structural properties of networks produced with different filtering methods is not straightforward. First, since the networks are dynamic, one needs to compare the structural properties of the static snapshots of the networks produced by the two methods at the same time. Second, parameters of the methods, i.e., forgetting factor $C_\text{f}$ and time-window width $T_\text{tw}$, influence the algorithms, so one needs to draw an equivalency between them to compare the methods under the same conditions. A natural condition to consider is the one of equal areas under the curves from Figure~2, representing the contribution of an interaction event to the representation of a node over time. Note that under this condition a node with constant non-zero activity in time will have the same strength in networks created with each method. For \texttt{fastviz}, the area $A_\text{fv}$ under the aggregation curve is equal to the sum of a geometric progression. Assuming an infinite geometric progression, we get the approximate $A_\text{fv} = T_\text{f}/(1-C_\text{f})$. The area under the aggregation curve of the rectangular time-window is simply $A_\text{tw}=T_\text{tw}$. By demanding the areas to be equal, we obtain the relation between the parameters of the two methods
\begin{eqnarray}
T_\text{tw} = \frac{T_\text{f}}{1-C_\text{f}}. \label{eq:params}
\end{eqnarray}
In general, the forgetting period $T_\text{f}$ is fixed, therefore there is only one free parameter controlling the filtering, e.g., the forgetting factor $C_\text{f}$, which we assign according to the dynamic network, i.e., the faster the network densifies in time, the more aggressive forgetting we use (see Appendix B for more details about the values of parameters).
In the following paragraphs, we analyze the dynamics of several structural properties of the networks produced with \texttt{fastviz}, rectangular, and exponential sliding time-window methods having equal aggregating areas.

To highlight the differences between the three filtering methods, we apply them to two real dynamic networks from Twitter characterized by high changeability and measure the structural properties of resulting networks (Figure~3). The networks represent interactions in Twitter during two widely popular events: the 2013 Super Bowl and the announcement of Osama bin Laden's death. Further description and properties of these datasets are provided in the next section.

\begin{figure*}[tbp]
\begin{center}
\includegraphics[width=17cm]{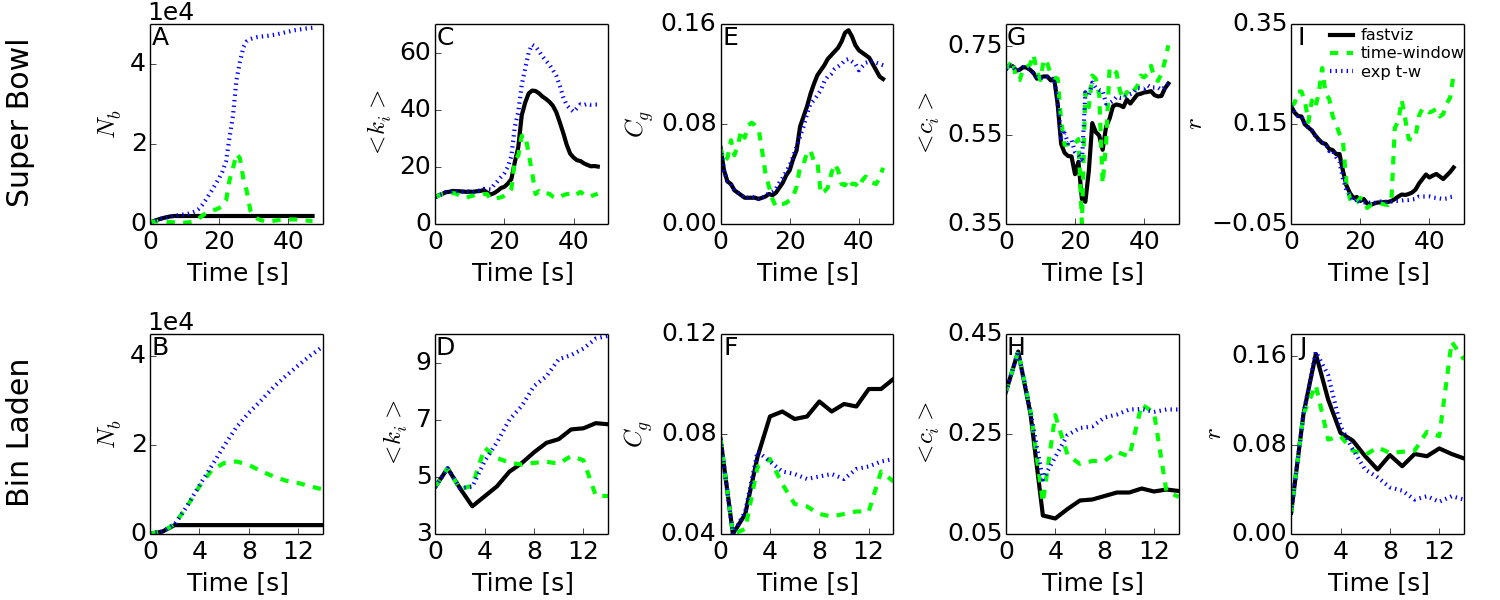}
\caption{Structural properties of filtered dynamic networks representing user interactions surrounding the 2013 Super Bowl or the announcement of Osama bin Laden's death. The values of the properties are plotted as a function of time for the \texttt{fastviz} filtering (black line), rectangular sliding time-window of matching width (green dashed), or exponential sliding time-window (blue dotted). The following network properties are plotted from the left-most to the right-most column: the total number of nodes $N_\text{b}$, the average degree $\left\langle k_i \right\rangle$, the global clustering coefficient $C_\text{g}$, the average local clustering coefficient $\left\langle c_i \right\rangle$, and the degree assortativity~$r$.}
\end{center}
\end{figure*}

First, we observe that, while the \texttt{fastviz} algorithm stores a limited number of nodes, i.e., up to $N_b=2000$ nodes, the sliding time-window methods store a varying number of nodes peaking at $20$ times more nodes than our method (Figures~3A and 3B). Due to this fact the computational complexity of sliding-time window methods increases in time, whereas it is bounded in \texttt{fastviz}. Since network structural properties such as average degree and clustering depend on the size of the network, we calculate these properties for the subgraphs of equal size, i.e., for the $N_b$ strongest nodes of the full network produced by each of the sliding time-window methods (Figures~3C-J). For simplicity, we refer to these subgraphs of $N_b$ nodes as the buffered networks.

Second, we find that the networks produced with our filtering method do not experience drastic fluctuations of the global and local clustering coefficients and degree assortativity, which are especially evident for the rectangular time-window (Figures~3E, 3G, 3H, and 3I). We conclude that the \texttt{fastviz} filtering produces smoother transitions between network snapshots than rectangular sliding time-window. This property of our method may improve readability of visualizations of such dynamic networks.

Finally, \texttt{fastviz} captures persistent trends in the values of the properties by leveraging the short-term and long-term node activity. For instance, it captures the trends in degree, clustering coefficients, and assortativity that are less visible with the rectangular time-window, while they are well-visible with the exponential time-window (Figures~3C-F, 3I, and 3J).
Note that high average degree obtained for networks produced with exponential time-window corresponds to the nodes that are active over a prolonged time-span, whose activity is aggregated over unbounded aggregation period, and the number of nodes is unbounded as well. On the contrary, rectangular sliding time-window shows the degree aggregated over a finite time-window, while \texttt{fastviz} limits the number of tracked nodes, leading to lower reported average degree.

\begin{figure}[tbp]
\begin{center}
\includegraphics[width=8.5cm]{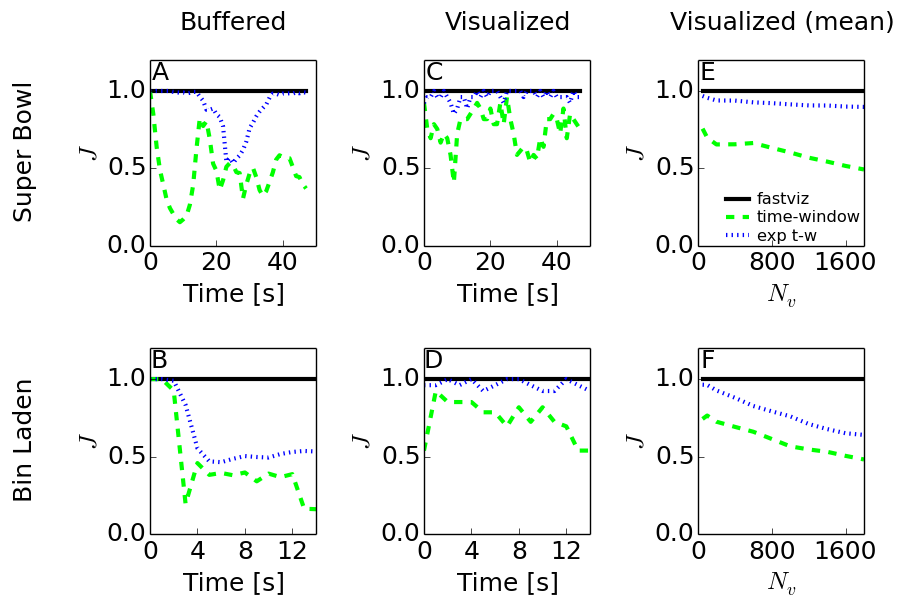}
\caption{Jaccard similarity coefficient $J$ between sets of nodes obtained with \texttt{fastviz} and rectangular sliding time-window (green dashed) or exponential sliding time-window (blue dotted). The nodes belong to either buffered or visualized networks representing Twitter interactions during the Super Bowl or Osama bin Laden's death. Specifically: (A-D) the Jaccard coefficient as a function of time; (E-F) the Jaccard coefficient averaged over time as a function of the number of nodes in the visualized network.}
\end{center}
\end{figure}

To measure the similarity of sets of nodes filtered with different methods we calculate Jaccard similarity coefficient. Specifically, we measure the Jaccard coefficient $J$ of the sets of $N_b$ strongest nodes filtered with \texttt{fastviz} and each of the time-window methods (Figures~4A and 4B). The value of the coefficient varies in time and among datasets. However, the similarity between \texttt{fastviz} and exponential time-window is significantly higher than between \texttt{fastviz} and rectangular time-window. For the Superbowl dataset, the similarity between \texttt{fastviz} and exponential time-window is close to $1$ most of the time and has a drop in the middle. The drop corresponds to the period of the game characterized by the intense turnout of nodes and edges in the buffered network. Hence, the similarity is not equal to $1$ for the two methods because the weakest nodes are often forgotten and interchanged with new incoming nodes in \texttt{fastviz}, while in exponential time-window method they are not forgotten and can slowly become stronger over time. In the next subsection we show that this similarity is close to $1$ at all times for the subsets of strongest nodes selected for visualization.

\subsection*{Network updates for visualization}

\begin{figure*}[tbp]
\begin{center}
\includegraphics[width=17cm]{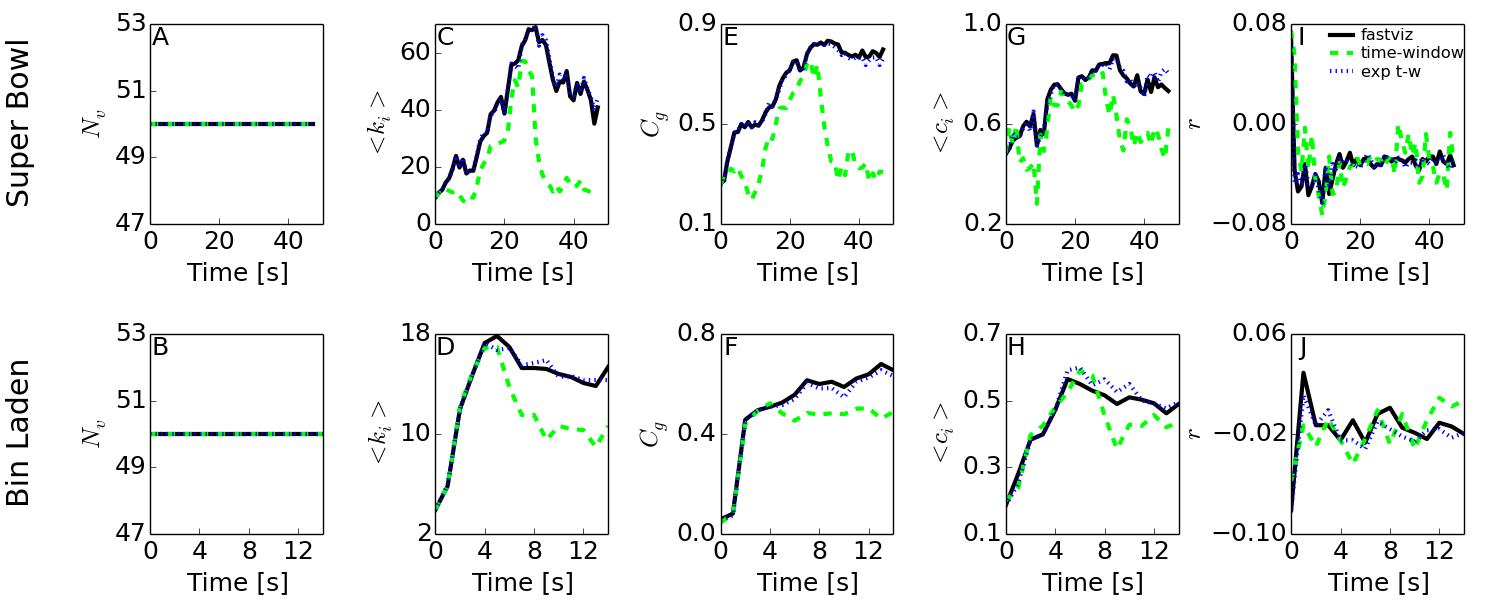}
\caption{Structural properties of the visualized dynamic networks representing Twitter interactions during the Super Bowl or Osama bin Laden's death. The values of the properties are plotted as a function of time for the \texttt{fastviz} filtering (black line), rectangular sliding time-window (green dashed), or exponential sliding time-window (blue dotted). The following network properties are plotted from the left-most to the right-most column: the total number of visualized nodes $N_\text{v}$, the average degree $\left\langle k_i \right\rangle$, the global clustering coefficient $C_\text{g}$, the average local clustering coefficient $\left\langle c_i \right\rangle$, and the degree assortativity~$r$.}
\end{center}
\end{figure*}

In the second stage, for the purpose of visualization, the algorithm selects $N_\text{v}<N_\text{b}$ nodes with the highest strength and creates a differential update to the visualized network consisting of these nodes and the connections between them. Each such differential update is meant to be visualized in the resulting animation of the network, e.g., as a frame of a movie.

We compare the visualized networks generated by each of the filtering methods. Each of the visualized networks consists of $N_v=50$ strongest nodes and all connections existing between them in the buffered network. The similarity of the nodes visualized by the \texttt{fastviz} and exponential time-window methods, measured as Jaccard coefficient $J$, is $1$ or close to $1$ (Figures~4C and 4D). The visualized networks of the two methods are almost identical. The structural properties of the networks created with the two methods yield almost the same values at each point in time (Figures~5A-J). This result is to be expected, since the forgetting mechanism of \texttt{fastviz} corresponds closely to the exponential decay of connection weights. The advantage of our method over exponential time-window consists of the limited computational complexity, which makes the \texttt{fastviz} filtering feasible even for the largest datasets of pairwise interactions. Naturally, the similarity between visualized networks created with the two methods decreases with the size of the visualized network $N_v$ (Figures~4E and 4F). More specifically, the similarity decreases with the ratio $N_v/N_b$, as we keep in our experiments a constant value of $N_b=2000$. Hence, to visualize larger networks one can choose to buffer more nodes.

The comparison of the evolution of structural properties of the corresponding buffered and visualized networks shows that these networks differ significantly for each of the filtering methods (compare Figure~3 vs. Figure~5). This difference is the most salient in the case of rectangular time-window, which yields considerably larger fluctuations of structural properties than the other methods. In the cases of \texttt{fastviz} and exponential time-window some structural properties show evolution that is qualitatively similar for buffered and visualized networks, e.g., the average degree and the global clustering coefficient (Figures~3C-F vs. Figures~5C-F). We conclude that the structure of visualized network differs significantly from the structure of buffered network, although this difference is smaller for \texttt{fastviz} than for rectangular sliding time-window.

\subsection*{Computational complexity}

The computational complexity of the buffering stage of the algorithm is $\mathcal{O}(E N_\text{b})$, where $E$ is the total number of the pairwise interactions read (the cliques are made of multiple pairwise interactions). Each time when an interaction includes a node that is not yet stored in the buffered graph the adjacency matrix of the graph needs to be updated. Specifically, the weakest node is replaced with the new node, so $N_b$ entries in the adjacency matrix are zeroed, which corresponds to $\mathcal{O}(E N_\text{b})$. The memory usage scales as $\mathcal{O}(N_\text{b}^2)$, accounting for the adjacency matrix of the buffered graph.\footnote{For certain real dynamic networks, the buffered graph is sparse. In such cases, one can propose more optimized implementations of \texttt{fastviz}. Here, we focus on limiting the time complexity so that it scales linearly with the number of interactions and describe the generic implementation that achieves it.}
The second, update-generating, stage has computational complexity of $\mathcal{O}(U N_\text{b}\text{log}(N_\text{b}))$, where $U$ is the total number of differential updates, which is a fraction of $E$ and commonly it is many times smaller than $E$.\footnote{Typically, a large number of interactions is aggregated to create one differential update to the visualized network. In the examples that we show in the next section, one update aggregates from $400$ to $2$ million interactions. Therefore, $U$ is from $400$ to $2$ million times smaller than $E$.} This term corresponds to the fact that the strengths of all buffered nodes are sorted each time an update to the visualized network is prepared. The memory trace of this stage is very low and scales as $\mathcal{O}(N_\text{v})$.
We conclude that our method has computational complexity that scales linearly with the number of interactions. It is therefore fast, that is, able to deal with extremely large dynamic networks efficiently.

\section{Visualization}

In this section, we describe animations of exemplary dynamic graphs filtered with \texttt{fastviz}.
Principally, the sequence of graph updates can be converted into image frames that are combined into a movie depicting the network evolution. We implement this visualizing technique and create with it the network animations described below. Alternatively, the updates can be fed directly to the Gephi Streaming API to produce an interactive visualization of the evolving network. The Gephi Streaming API allows graph streaming from a client to a server where Gephi is running. In such a case, the graphs are streamed directly from our filtering system to the Gephi server without any third-party modules.
In Appendix A, we introduce implementation details of both approaches. Finally, corresponding animations can be created by other visualization tools fed with the \texttt{fastviz} updates; we highly encourage their development.

\subsection*{Datasets}

We test the \texttt{fastviz} filtering and our visualizing technique on four datasets very different from each other in nature, size, and time span (see Table~1). The datasets and movies produced from each dataset are described in the following subsections (see Figure~6). In Appendix A, we present the source code of both tools with their documentation, four dynamic graph datasets, and instructions to recreate the visualizations introduced in this section. In Appendix B, we provide and describe the values of the parameters of the algorithm and the visualizing tool used for these datasets.

\begin{figure}[tbp]
\centerline{\includegraphics[width=8.5cm]{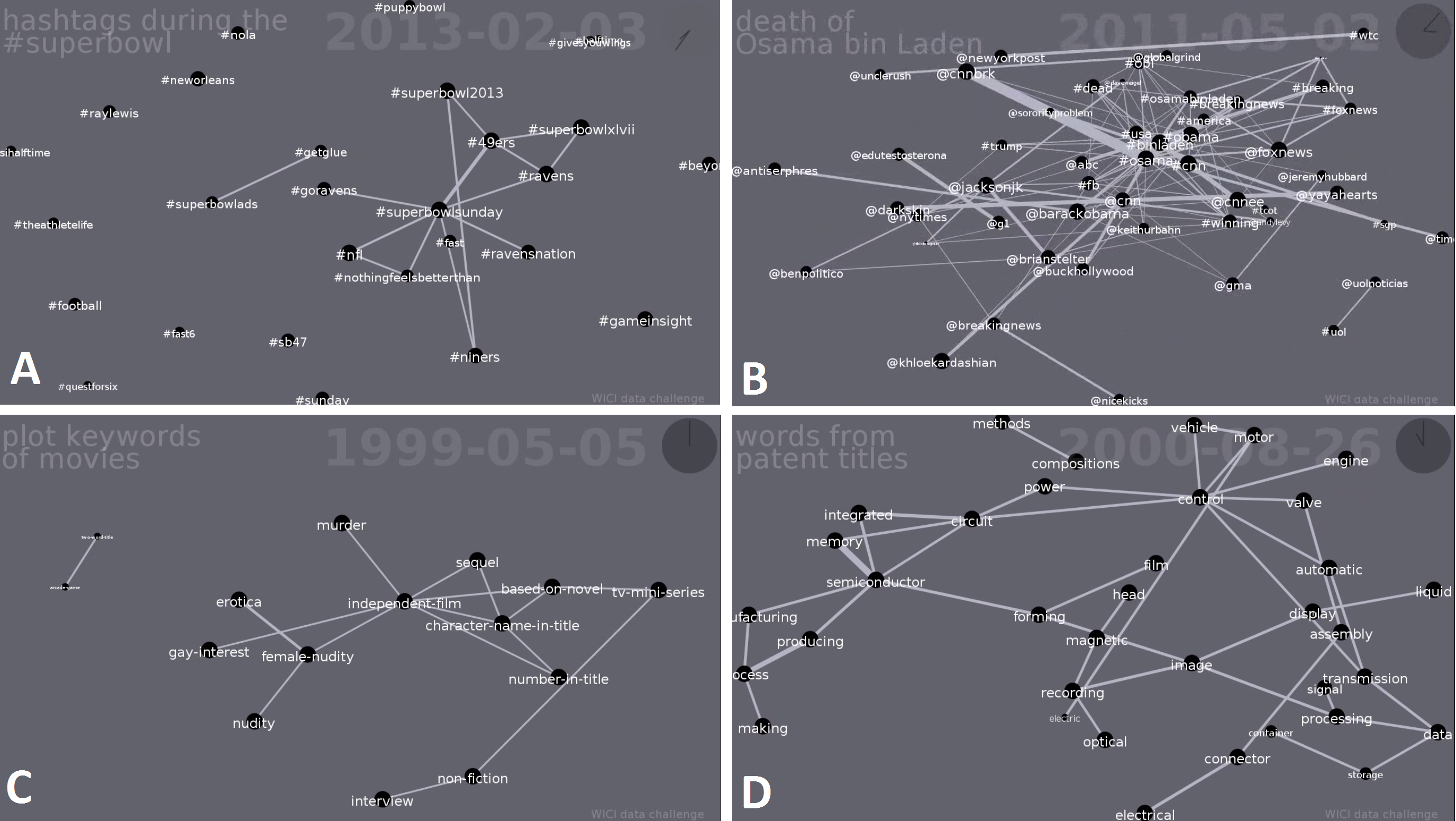}}
 \caption{Screenshots of the movies generated from the datasets: (A) Super Bowl (full animation is available at \url{http://youtu.be/N1wmJG3dVhs}), (B) Osama bin Laden's death (\url{http://youtu.be/gk03CJDAp_w}), (C) IMDB keywords (\url{http://youtu.be/JxWGjMdLUdQ}), (D) US Patents (\url{http://youtu.be/Q7p-bRY7_n0}).}
\label{fig:snapshots}
\end{figure}

\subsection*{Twitter}

We use data obtained through the Twitter \textit{gardenhose} streaming API, which covers around $10$\% of the tweet volume. We focus on two events: the announcement of Osama bin Laden's death and the 2013 Super Bowl. We consider user mentions and hashtags as entities and their co-occurrence in the same tweet as interactions between them.

The first video (Figure~2A) shows how the anticipation for the Super Bowl steadily grows on early Sunday morning and afternoon, and how it explodes when the game is about to start. Hashtags related to \#commercials and concerts (e.g., \#beyonce) are evident. Later, the impact of the \#blackout is clearly visible. The interest about the event drops rapidly after the game is over and stays low during the next day.

The video about the announcement of Osama bin Laden's death (Figure~2B) shows the initial burst caused by @keithurbahn and how the breaking news was spread by users @brianstelter and @jacksonjk. The video shows that the news appears later via \#cnn and is announced by @obama. The breaking of this event on Twitter is described in detail by Lotan~\cite{binladen}.

\subsection*{IMDB movies}

We use a dataset from IMDB of all movies, their year of release and all the keywords assigned to them (from \url{imdb.to/11SZD}). We create a network of keywords that are assigned to the same movies. Our video (Figure~2C) shows interesting evolution of the keywords from ``character-name-in-title'' and ``based-on-novel'' (first half of 20th century), through ``martial-arts'' (70s and 80s) to ``independent-film'' (90s and later), ``anime'' and ``surrealism'' (2000s).

\subsection*{Patents}

We use a set of US patents issued between $1976$ and $2010$~\cite{Rowe2009Scholarly}. We analyze the appearance of words in their titles. Whenever two or more words appear in a title of a patent we create a link between them at the moment when the patent was issued. To improve readability we filter out stopwords and the generic frequent words: ``method,'' ``device'' and ``apparatus.'' Our video (Figure~2D) shows that at the beginning of the period techniques related to ``engine'' and ``combustion'' were popular, and later start to cluster together with ``motor'' and ``vehicle.''  Another cluster is sparked by patents about ``magnetic'' ``recording'' and ``image'' ``processing.'' It merges with a cluster of words related to ``semiconductor'' and ``liquid'' ``crystal'' to form the largest cluster of connected keywords at the end of the period.

\subsection*{Other visualizations}
Other than these experimental datasets, on-demand animations of Twitter hashtag co-occurrence and diffusion (retweet and mention) networks can be generated with our tool via the Truthy service (\url{truthy.indiana.edu/movies}). Hundreds of videos have already been  generated by the users of the platform and are available to view on YouTube (\url{youtube.com/user/truthyatindiana/videos}).

\subsection*{Summary}

The datasets in our case studies are fairly diverse in topicality, time span, and size, as shown in Table~1. Nevertheless, our method is able to narrow down the visualization to meaningful small subgraphs with less than $600$ distinct nodes in all cases.
The high performance of the algorithm makes it viable for real-time visualizations of live and large data streams. On a desktop machine the algorithm producing differential updates of the network took several minutes to finish for the US patents and less than two minutes for the other datasets. Given such a performance, it is possible to visualize in real-time highly popular events such as the Super Bowl, which produced up to $4,500$ tweets per second.

\section{Conclusions}


Tools for dynamic graph visualization developed so far do not provide specialized ways to dynamically select the most important portions of large evolving graphs. We contribute to filling this gap by proposing an algorithm to filter nodes and edges that best represent the network structure at a given time. Our method captures trends and smoothens the dynamics of structural properties of weighted networks by leveraging the short-term and long-term node activity. Furthermore, our filtering method uses limited memory and processor time making it viable for large live data streams.
We implemented our filtering algorithm in open source tools that take in input a stream of interaction data and output a movie of the network evolution or a live Gephi animation.
As future work, we wish to improve our algorithm by means of further optimization and to enhance the tools by providing a standalone module for live visualization of graph evolution.

\bigskip

\section*{Author's contributions}
All authors designed the research. PAG wrote the source code of the algorithm and LMA wrote the source code of the visualization tool. All authors deployed the tools. PAG and LMA analyzed the data. All authors wrote, reviewed and approved the manuscript.

\section*{Acknowledgements}

We are grateful to Andr\'{e} Panisson for inspiration and to Jacob Ratkiewicz, Bruno Gon\c{c}alves, Mark Meiss, and other members of the Truthy project (\url{cnets.indiana.edu/groups/nan/truthy}) for helpful discussions and suggestions. PAG acknowledges funding from the JAE-Predoc program of CSIC and partial financial support from the MINECO under project MODASS (FIS2011-24785). This work is supported in part by the NSF (ICES award CCF-1101743), and the James S. McDonnell Foundation.

\bibliographystyle{abbrv}  
\bibliography{epjds}

\appendix
\section{Implementation details and \\source code}

We have implemented two independent tools described in the manuscript. The first tool is the \texttt{fastviz} algorithm. The second tool converts the sequence of updates into image frames that are combined into a movie depicting the network evolution. We release the source code of both tools (see the project website \url{github.com/WICI/fastviz}). Here, we describe the two tools in more detail. 

The first tool is the \texttt{fastviz} algorithm. It takes in input a chronological stream of interactions between nodes and converts it into a set of graph updates that account only for the most relevant part of the network in the JSON format. 
In the network filtering stage, the algorithm stores a buffered network of size $N_b$, limiting the computational complexity and memory usage of the algorithm.
In the second stage, for the purpose of visualization, the algorithm selects $N_\text{v}<N_\text{b}$ nodes with the highest strength and all edges between these nodes that have weight above a certain threshold $w^\text{min}$. The subgraph induced by the $N_\text{v}$ nodes is compared with the subgraph in the previous state and a differential update is created. The updates are created per every time interval that is determined with the time contraction parameter $T_\text{c}$. A value of $10$ for this parameter means that the time will flow in the visualization $10$ times faster than in the data given as the input (see Appendix B).
The differential updates are written in output in the form of a JSON file formatted according to the Gephi Streaming API \cite{graphstreamingapi}. We choose JSON format specifically due to the compatibility with Gephi Streaming API. In short, each line of the JSON file corresponds to one update of the graph structure and contains a sequence of JSON objects that specify the addition/deletion/attribute change of nodes and edges. We also introduced a new type of object to deal with labels on the screen, for example, to write the date and time on the screen.

The second tool converts the sequence of updates into image frames that are combined into a movie depicting the network evolution.
To this end, the sequence of updates produced by the filtering algorithm is fed to a python module that builds a representation of a \textit{dynamic graph}, namely an object that handles each of the updates and reflects the changes to its current structure. The transition between the structural states of the graph determined by the received updates is depicted by a sequence of image frames. Each differential update correspond to one visualization frame, i.e., one frame of an animation. In its initial state, the nodes in the network are arranged according to the Fruchterman Reingold graph layout algorithm~\cite{fruchterman91graph}. The choice of the layout is arbitrary and other layouts can be used and compared. However, due to the focus of this study on the filtering method, rather than the quality of the visualization, we do not explore any other layout algorithms. For each new incoming event, a new layout is computed by running $N$ iterations of the layout algorithm, using the previous layout as a seed. Intermediate layouts are produced at each iteration of the algorithm. Every intermediate layout is converted to a png frame that is combined through the \emph{mencoder} tool \cite{mencoder} to produce a movie that shows a smooth transition between different states. The movie is encoded with the frequency of $30$ frames per second. To avoid nodes and edges to appear or disappear abruptly in the movie, we use animations that smoothly collapse dying nodes and expand new ones. A configuration file allows to modify the default movie appearance (e.g, resolution, colors) and layout parameters (see the project website).

We release the source code of both tools with the documentation under the GNU General Public License (see the project website \url{github.com/WICI/fastviz}). Together with the tools we release the datasets used in this paper and instructions on how to recreate all the examples of animations presented in this manuscript.
Additionally, the updates created with \texttt{fastviz} can be fed directly to the Gephi Streaming API to produce an interactive visualization of the evolving network. Respective instructions can be found at the website of the project.

\section{Algorithm parameters}

The exact behavior of the \texttt{fastviz} filtering depends on the parameters introduced in the manuscript. We present the values of the parameters used in the case studies and their default values in Table~2.
The default values of the parameters are meant to be universal and give reasonably good visualizations for most datasets.
Overall, three parameters require adjustment to the input data, namely time contraction $T_\text{c}$, edge width threshold $w^\text{min}$, and forgetting factor $C_\text{f}$. We provide exemplary values of these parameters for the introduced datasets in Table~2 and describe these parameters in detail below.

The time contraction $T_\text{c}$ corresponds to the number of seconds in data time scale that are going to be contracted to one second of the visualization. The larger the time span of the dataset, the larger should be this parameter in order to keep the length of visualization fixed. For instance, if the timespan of the network is $10$ hours, and one wants to see its evolution in a $10$-second-long animation, then $T_c$ should be set to $3600$. It is crucial to provide a desired value for this parameter, because providing a value that is too large will create just a few network updates and a very short animation, while providing a value that is too small will create a large number of updates making the JSON file very big and the animation very long.

The minimal edge weight $w^\text{min}$ is a threshold above which edges appear in the visualization. Low value of this parameter may results in many edges of low weight appearing in the animation, while high value of the parameter may prevent any edges from being visualized. In case a user does not have any information about the visualized network, we recommend leaving this parameter at its default value of $0.95$, which will visualize all edges of standard weight $1$ or higher.

The forgetting factor $C_\text{f}$ decides how fast older interactions among nodes are forgotten in comparison with more recent interactions. This parameter can be tuned individually for the purpose of the visualization. In general, the faster the network densifies in time, the more aggressive should be the forgetting, i.e., the lower should be the forgetting factor $C_\text{f}$. In general, keeping the default value of this parameter is safe, although its adjustment will improve the quality of visualization.

\end{document}